\documentclass[aps,pra,reprint,showpacs,superscriptaddress,nofootinbib]{revtex4-2}

\usepackage{graphicx}
\usepackage{dcolumn}
\usepackage{bm}
\usepackage{color}
\usepackage{scrextend}
\usepackage{url}
\usepackage{epstopdf}
\usepackage{hyperref}
\usepackage{courier}
\usepackage{todonotes}
\usepackage{amssymb}
\usepackage{censor}
\usepackage{soul}  

\usepackage{xcolor}
\definecolor{removecolor}{RGB}{219, 48, 122}
\definecolor{appendixcolor}{RGB}{48, 219, 219}
\definecolor{suggest}{RGB}{0, 000, 250}

\newcommand{\toappendix}[1]{#1}

\begin{document}
\StopCensoring

\title{Numerical simulation projects in micromagnetics with Jupyter}

\censorbox{  
\author{Martin Lonsky}
\email{lonskymartin@gmail.com}
\affiliation{Materials Research Laboratory and Department of Materials Science and Engineering, University of Illinois at Urbana-Champaign, Urbana, Illinois 61801, USA}
\affiliation{Institute of Physics, Goethe University Frankfurt, 60438 Frankfurt, Germany}
\author{Martin Lang}
\affiliation{Max Planck Institute for the Structure and Dynamics of Matter,
  22761 Hamburg, Germany}
\affiliation{Faculty of Engineering and Physical Sciences, University of Southampton, Southampton, SO17 1BJ, United Kingdom}
\author{Samuel Holt}
\affiliation{Faculty of Engineering and Physical Sciences, University of Southampton, Southampton, SO17 1BJ, United Kingdom}
\affiliation{Max Planck Institute for the Structure and Dynamics of Matter, 22761 Hamburg, Germany}
\author{Swapneel Amit Pathak}
\affiliation{Faculty of Engineering and Physical Sciences, University of Southampton, Southampton, SO17 1BJ, United Kingdom}
\affiliation{Max Planck Institute for the Structure and Dynamics of Matter, 22761 Hamburg, Germany}
\author{Robin Klause}
\affiliation{Materials Research Laboratory and Department of Materials Science and Engineering, University of Illinois at Urbana-Champaign, Urbana, Illinois 61801, USA}
\author{Tzu-Hsiang Lo}
\affiliation{Materials Research Laboratory and Department of Materials Science and Engineering, University of Illinois at Urbana-Champaign, Urbana, Illinois 61801, USA}
\author{Marijan Beg}
\affiliation{Department of Earth Science and Engineering, Imperial College London, London SW7 2AZ, United Kingdom
}
\author{Axel Hoffmann}
\affiliation{Materials Research Laboratory and Department of Materials Science and Engineering, University of Illinois at Urbana-Champaign, Urbana, Illinois 61801, USA}
\author{Hans Fangohr}
\email{hans.fangohr@mpsd.mpg.de}
\affiliation{Max Planck Institute for the Structure and Dynamics of Matter, 22761 Hamburg, Germany}
\affiliation{Faculty of Engineering and Physical Sciences, University of Southampton, Southampton, SO17 1BJ, United Kingdom}
\affiliation{Center for Free-Electron Laser Science, 22761 Hamburg, Germany}
}

\date{\today}

\begin{abstract}
  We report a case study where an existing materials science course was
  modified to include numerical simulation projects on the micromagnetic behavior of materials. The Ubermag micromagnetic simulation software package is used in order to solve problems computationally. The simulation
  software is controlled through Python code in Jupyter notebooks. 
  Our experience is that the self-paced problem-solving nature of the project
  work can facilitate a better in-depth exploration of the course
  contents. We discuss which aspects of the Ubermag and the
  project Jupyter ecosystem have been beneficial for the students' learning experience
  and which could be transferred to similar teaching activities in other subject
  areas.
\end{abstract}

\maketitle

\section{Introduction} \label{intro} Traditionally, science curricula at the university level consist of theory-focused classes and experimental courses. However, not only in the natural sciences, but also within the engineering community, computation has emerged as a third fundamental methodology~\cite{Skuse_2019}. Both experimentalists and
theorists make use of computational techniques in their activities.
Oftentimes, a system of interest is too complex to be solved analytically or
certain experiments cannot be carried out in a laboratory. In such cases,
numerical studies can help to improve our understanding.

In STEM education, computational modeling has an important
role~\cite{Weber_2020}, and it is widely argued that more computational content
in curricula would be desirable~\cite[{\em e.g.},][]{Caballero_2018}.
Anecdotal evidence on
undergraduate programs at numerous universities worldwide suggests that
computational contents remain severely underrepresented in the respective
curricula~\cite{Caballero_2018, Kortemeyer_2020},
despite recent studies presenting evidence that computational methods education
in undergraduate coursework may lead to the development of multiple essential
skills~\cite{Graves_2020, Cook_2008}. For example, the American
Association of Physics Teachers (AAPT) has identified competency in computation
to be vital for success at the workplace or PhD research activities for
physicists~\cite{McNeil_2017, AAPT_2016}.
Computational modeling and numerical simulations appear crucial for obtaining a complete
picture of the modern STEM disciplines, and therefore adequate ways need to be
found to embed this branch into teaching curricula.

We can distinguish between at least two approaches toward
incorporating computational methods in a curriculum~\cite{Chonacky_2008}: Firstly, there are courses
which solely focus on programming, numerical methods, modeling, and
simulations. Secondly, computational
content can also be introduced by embedding it in existing courses. The latter
approach is at the core of this case study.

Here we report on the introduction and implementation of numerical simulation
group projects in an elective course within the materials science and
engineering curriculum of the University of \censor{Illinois at
Urbana-Champaign (UIUC)}. The course is also available to students in other fields, such as electrical and computer engineering or physics, and requires basic knowledge in condensed matter physics. The students set up and perform
micromagnetic simulations by using the open computational
environment\footnote{Open computational environments allow students to directly
  see and control the underlying algorithm of the computational model, while
  closed computational environments such as simulation applets are considered as
  a black box with no or little information about the exact model~\cite{Caballero_2012}.}
Ubermag~\cite{Beg_2022}. In this article, we describe our experiences using Ubermag and related computational software packages in STEM instruction. We discuss our insights from the teaching delivery, student
evaluations and personal interview surveys.

The paper is structured as follows.
Section \ref{Sec3} contains a description of the course on magnetic materials and applications at \censor{UIUC}, a brief introduction to computational micromagnetics, and a
detailed presentation of the Ubermag software and
its application in the classroom.
Based on the students' feedback and our own experience, we give recommendations for the implementation of computational projects in other courses in Sec.~\ref{sec:recommendations}. We provide supplementary material along with this article, including a general overview on the incorporation of computational contents into STEM programs, a description of simulation projects and the corresponding problem sheets, additional practical considerations, and a thematic analysis of the feedback from students and the teaching staff.  

\section{Micromagnetic Simulation Projects in a Materials Science and Engineering Course} \label{Sec3}\label{sec:micromagnetic-simulation-project}

There exist many approaches to integrating computational contents into
undergraduate or graduate STEM degree curricula (see Sec.~I of supplementary material). Here, we present a case study that we conducted within the framework of a class
on ``Magnetic Materials and Applications''. We utilized the software package
Ubermag~\cite{Beg_2022}, developed at the University of Southampton, United
Kingdom, and the Max Planck Institute for the Structure and Dynamics of Matter,
Germany, which provides a Python interface to existing micromagnetic simulation
packages.

We introduced group projects that make use of the Ubermag software package. Ubermag offers an easy-to-learn approach to create, control and run simulation scripts that
solve the underlying partial differential equation that describes the temporal
evolution of the magnetic field in a specified materials system.

In Subsec.~\ref{sec:elect-course-magn}, we begin with a detailed description of
the Magnetic Materials and Applications course in which we have conducted our case study. This is followed by an introduction to (analytical) micromagnetic theory in Subsec.~\ref{sec:intr-micr} and the numerical computation of solutions in Subsec.~\ref{sec:intr-comp-micr}. 
Finally, Subsec~\ref{sec:ubermag} introduces the Ubermag software.

\subsection{The Elective Course on Magnetic Materials and Applications} \label{sec:elect-course-magn}

The Magnetic Materials and
Applications class \censor{(MSE 598/498/464) at UIUC} is an elective course aimed at
both undergraduate and graduate students at the Department of Materials Science
and Engineering, but other students from the physics, chemical engineering
and electrical engineering departments have also attended this class. The total enrollment ranges from 7 to 15 students per semester.

The lecture introduces the fundamental concepts with regard to the practical use of
magnetic materials. The course objectives are: 
\begin{itemize}
\item Understand how different magnetic interactions determine static and dynamic magnetic properties.
\item Quantify essential magnetic materials properties.
\item Design components of magnetism-based devices.
\item Use basic micromagnetic simulations.
\end{itemize}

The class is held over the span of about 16 weeks
and it is recommended that students dedicate 6-8 hours per week
to working on the course. Aside from the live lectures, online discussions are
encouraged via Canvas (an online course and learning management system), weekly homework is assigned, literature review
presentations are delivered by the students, and a micromagnetic simulation
project has to be completed successfully.

\begin{figure*}
  \centering
  \includegraphics[width=1\textwidth]{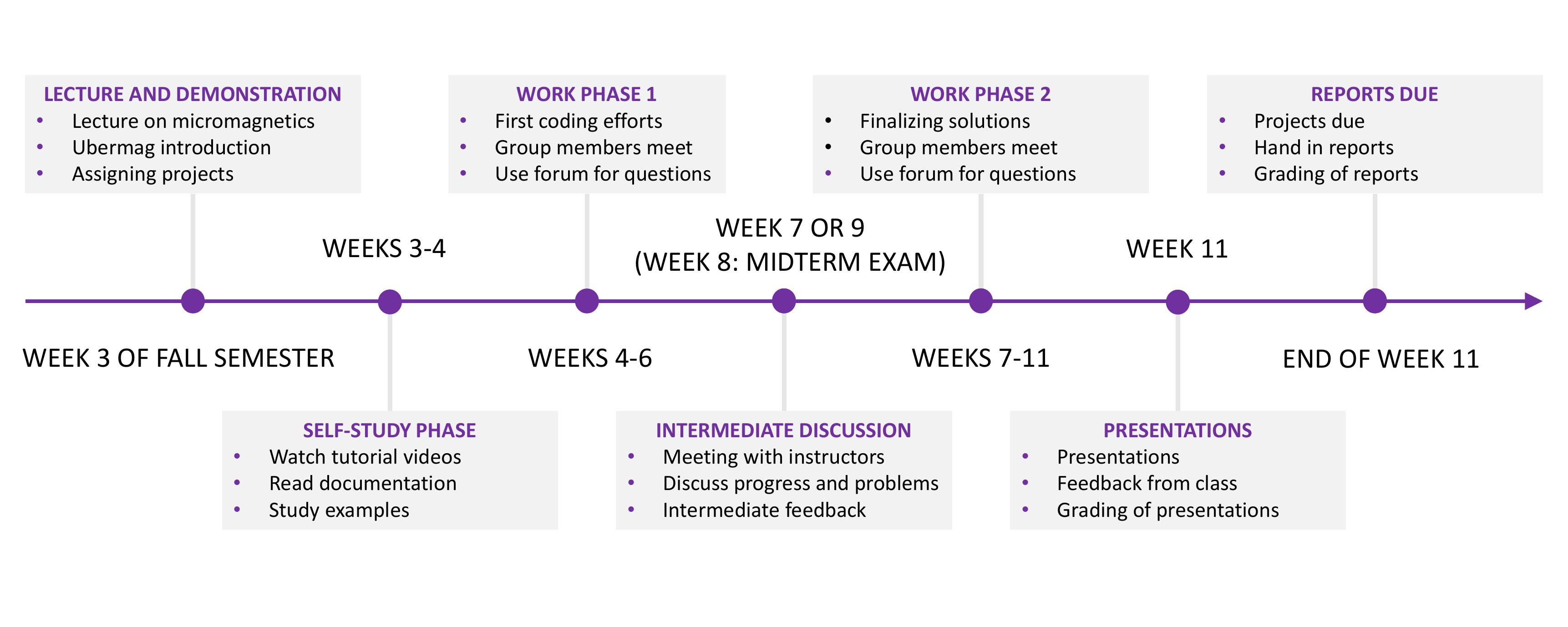}
  \caption{Example timeline for the computational micromagnetics projects.} 
  \label{TIMELINE_PROJECTS}%
\end{figure*}%

We have designed five distinct simulation projects. Students are asked to
work in groups of two to four, since each project is divided into several
subprojects, that are mostly independent of each other, but do have a
certain overlap such that it is beneficial for the students to interact with
their peers and discuss their solutions. More detailed information about the contents of the projects and problem sheets can be found in Sec.~II of the supplementary material. A sample timeline of the
projects is illustrated in
Fig.~\ref{TIMELINE_PROJECTS}. Before the problems are handed out to the
students, we give a brief introduction (about 30 minutes) to Ubermag as part of a standard 90-minute class that focuses on micromagnetic simulations. Furthermore, we provide them with additional material such
as video tutorials by the Ubermag developers and the accompanying
software documentation, which is very comprehensive and includes numerous
examples of Jupyter notebooks that enable to run Ubermag. Due to the students' diverse backgrounds, their
exposure to programming in general and Python in particular prior to working on
the computational micromagnetics projects has been vastly different. 
For instance,
the Department of Materials Science and Engineering at \censor{UIUC} has computational
methods embedded in several core classes of the curriculum~\cite{Kononov_2017,
  Mansbach_2016, Mansbach_2016b, Zhang_2018, Lee_2019}, while students from
other majors who attend our course
may never have written their own code. Furthermore, it is reasonable to assume a
discrepancy in the average computational literacy between undergraduate and
graduate students in the class. 

Project reports are due around two months after the projects have been
assigned. It is prudent to set up
meetings between students and the instructor together with a teaching assistant
halfway through the duration of the computational project. Firstly, this
provides preliminary feedback to the
students and helps to prevent them from getting lost in details. Furthermore, it also enables
students to ask questions about the subject matter, programming in general, and
the instructor's expectation with regard to their report and presentation.
Lastly, it may be perceived as an intermediate deadline and thereby encourages
students to get started with the projects as early as possible. We also ensure
that students always have the possibility of reaching out to the teaching
assistants via email as well as on a Canvas discussion forum. A few weeks after the
intermediate discussions, students are required to present their results to the
class and then hand in a project report a few days later. After each presentation, we
aim to stimulate a technical discussion and then solicit feedback from the
audience on the presentation content and style.

\subsection{Introduction to Micromagnetics}\label{sec:intr-micr}
Micromagnetics is concerned with magnetization processes on length scales large enough to overlook atomic structure details of a material but small enough to resolve magnetic textures like domain walls. Examples of relevant applications include magnetic data storage devices and nanoparticles for medical purposes.     
The basis of time-dependent micromagnetics is the equation of motion of the
magnetization vector field (Landau-Lifshitz-Gilbert, LLG equation)~\cite{Gilbert_2004}:
\begin{equation} \label{eq:LLG}
  \frac{\mathrm{d}\textbf{m}}{\mathrm{d}t}=-|\gamma_{0}|\textbf{m}\times \textbf{H}_{\mathrm{eff}}(\textbf{m})+\alpha \textbf{m}\times \frac{\mathrm{d}\textbf{m}}{\mathrm{d}t}.
\end{equation} 
Here, $\gamma_{0}$ denotes the gyromagnetic constant, and $\alpha$ is the Gilbert
damping constant. The entity of interest is the magnetization vector field
$\textbf{m}(\textbf{r}, t) \in \mathbb{R}^3$ defined as a function of position
$\textbf{r} \in \mathbb{R}^3$ and time $t \in \mathbb{R}$. For a time-dependent problem, one
generally knows an initial magnetization vector field $\textbf{m}_0 =
\textbf{m}(t_0)$ at time $t_0$ and wants to compute $\textbf{m}(\textbf{r}, t)$
for $t>t_0$.

A significant part of the complexity originates from the effective field,
$\textbf{H}_\mathrm{eff}$, which is itself a function of the magnetization
vector field. The effective field can be computed from the energy $E$ of the system: 
\begin{eqnarray} \label{eq:Heff}
  \textbf{H}_\mathrm{eff} = -\frac{1}{\mu_0}\frac{\delta E}{\delta
  \textbf{m}}.
\end{eqnarray}

Different phenomena of material physics can be described by including different
contributions to the energy~$E$, for example
\begin{eqnarray} \label{eq:Utot}
  E(\textbf{m}) &=& E_{\mathrm{Ex}}(\textbf{m}) + E_{\mathrm{Z}}(\textbf{m}) +
                    E_{\mathrm{Dem}}(\textbf{m})\\\nonumber
  &&+ E_{\mathrm{Anis}}(\textbf{m}) + E_{\mathrm{DMI}}(\textbf{m}),
\end{eqnarray}
where $E_{\mathrm{Ex}}$ denotes the exchange energy, $E_{\mathrm{Z}}$ the Zeeman energy, $E_{\mathrm{Dem}}$ the
demagnetization energy, $E_{\mathrm{Anis}}$ the anisotropy energy, and
$E_{\mathrm{DMI}}$ the Dzyaloshinskii-Moriya interaction (DMI). All energy terms involve integrals over the volume, some involve vector
analysis operators, and the demagnetization term contains a
double integral over the volume due to the long-range nature of demagnetization
effects.

In summary, the micromagnetic problem --- summarized through Eqs.~(\ref{eq:LLG}) and
(\ref{eq:Utot}) --- is complex. Mathematically, this is reflected
in Eq.~(\ref{eq:LLG}) being a non-linear integro-partial differential equation. A rich variety
of phenomena are described by this model, ranging from dynamic effects such as ferromagnetic resonance and spin-wave propagation~\cite{Lonsky_2020a, Lonsky_2020, Lonsky_2022} to static equilibrium configurations of the magnetization field such as magnetic domains and vortices. It is this complexity and richness that makes the model a fruitful ground for advanced materials physics education.

\subsection{Introduction to Computational Micromagnetics}\label{sec:intr-comp-micr}

The micromagnetic model can only be solved analytically for a small number of cases
(often in geometries with particular symmetries). In general, a numerical
  approach is required to obtain a solution. A typical numerical approach toward the solution of the LLG
equation~(\ref{eq:LLG}) is given by
discretizing it in space using finite elements or finite differences. The
time-dependent problem then becomes numerically tractable by solving the spatial
partial differential equation for a time $t$, then advancing the solution from $t$ to $t + \Delta t$ through solving
a set of ordinary differential equations. We note that this iterative solution over steps in
time is algorithmically similar to the time integration method that is frequently used in VPython simulations. 

There are at least two widely used software packages that solve the complex
computational micromagnetics problem using finite differences and relying on the same physical model: the Object
Oriented MicroMagnetic Framework (OOMMF)~\cite{Donahue1999} and
mumax$^3$~\cite{Vansteenkiste_2014}. 
The OOMMF software operates on the computer's CPU, whereas mumax$^3$ is GPU-accelerated and
requires an Nvidia GPU card to be installed. OOMMF is
written in C++ and Tcl, mumax$^3$ is based on the programming languages Go
and CUDA. The input scripts for the simulations need to be defined by
the user in Tcl and a Go-like scripting language, respectively. The learning
curve for either package is long; while clearly acceptable in professional
research activities, it is a challenge for occasional users such as students in
an educational setting. In the remainder of this article, we will demonstrate that the Ubermag software has a significantly shorter learning curve, making it more suitable for educational use. Ubermag has been developed to offer a Python interface~\cite{Beg_2017} to
OOMMF with the goal of providing an improved environment for
researchers to support computational science investigations of magnetic
materials and devices. 
Later, Ubermag was extended to also interface with
mumax$^3$~\cite{Fangohr2024}. 
In what follows, we will provide more detailed information on Ubermag and how it can be used for teaching activities.

\subsection{The Ubermag Software and its Utilization in the Classroom}\label{sec:ubermag}
The Python packages provided by Ubermag allow the user to
specify micromagnetic models, run simulations, and analyze and visualize
data in interactive Jupyter notebooks, see Fig.~1 of the supplementary material. Only the computational solving of micromagnetic
problems is delegated to the \emph{micromagnetic calculators} ({\em i.e.}, OOMMF or
mumax$^3$), while all other steps are independent from these
simulation packages.

The Ubermag Python packages (Sec.~\ref{sec:conc-simul-driv}) are structured
to mirror the computational modelling concepts: define a physics model
(\texttt{micromagneticmodel}), discretise space (\texttt{discretisedfield}),
compute the numerical solution (\texttt{oommfc} and \texttt{mumax3c}).

Students can control and run their Ubermag simulations via
browser-based interactive Jupyter notebooks.
The modular structure of Jupyter notebooks allows running blocks of code (so
called ``cells'') individually instead of running the entire simulation script.
Students can obtain an in-depth
understanding of the underlying physics by iteratively modifying and exploring the
system (Sec.~\ref{sec:jupyter-notebook}).

The installation of software for teaching purposes can be challenging: The
university's or the students' personal laptops may be running a variety of
operating systems (typically Windows, MacOS or Linux) with different versions.
More complex simulation software environments may need multiple libraries of
compatible versions to be installed simultaneously. For the Ubermag software,
there are fortunately multiple ways to install it: Using conda-forge the three
main operating systems are supported. An installation using Python's standard
installation tool pip is also possible, but requires the user to manually
install a micromagnetic calculator (such as OOMMF or mumax$^3$).

All simulation projects in our course are carefully designed such that each
calculation runs for a reasonably short period of time, i.e., seconds to
minutes. For computational problems that can be computed within a few minutes on a
single-core CPU, there is another \emph{zero-install} way of using Ubermag
through a service called MyBinder available at \url{mybinder.org}.
In short: Ubermag can be executed in the cloud and controlled from any browser; no installation on the computer is necessary. This has been
very popular with students (see Sec.~\ref{sec:binder} for more details).

We discuss the value of using open source software in education in Sec.~III of the
supplementary material. Furthermore, we present a qualitative thematic analysis of the learning experience from the student and teacher perspective in Sec.~IV of the
supplementary material. In the following section, we offer suggestions for embedding computational projects into other courses based on the feedback and our experience.

\section{Recommendations for computational projects} \label{sec:recommendations}

The feedback we have obtained from students and teachers suggests
that computational problem solving can improve the
learning experience. In this section, we discuss the teaching
setup with the objective to extract insights that could be useful in
other subjects ({\em i.e.}, outside micromagnetics and materials physics more
generally). We want to comment on three points here: the choice of programming language
(\ref{sec:choice-of-programming-language}), the opportunities from the Jupyter
Notebook for use in education (\ref{sec:jupyter-notebook-for-education}), and
aspects of the Ubermag design that are beneficial for teaching
(\ref{sec:ubermag-design}).

\subsection{Choice of Programming Language} \label{sec:choice-of-programming-language}

The use of Python as the language to both drive the simulation
and to carry out the analysis of the data extracted from the simulations appears to be a good choice. Python is easy to learn yet very
powerful~\cite{Fangohr_2004}. Of particular relevance is the wide set of
Python libraries available for science and engineering --- including sophisticated data
analysis and data visualization tools. 

\subsection{Project Jupyter Tools for Education \label{sec:jupyter-notebook-for-education}}

\toappendix{The Jupyter notebook~\cite{Kluyver2016} emerged from the Interactive Python
(IPython)~\cite{Perez2007} environment. A recent review~\cite{Granger2021} by
the original authors makes the observation that the notebook has been designed to
help scientists think. As such, it is perhaps not suprising that the
Jupyter notebook has become the standard in data
science~\cite{Perkel2018},
and is increasingly used in science for data exploration and
analysis~\cite{Fangohr2020}. Students can benefit in similar ways as data scientists and scientists from
the Jupyter notebook, which is increasingly used in educational
settings~\cite{Barba2022,nbgrader2019}.
}

\subsubsection{Jupyter Notebook}\label{sec:jupyter-notebook}

\toappendix{The combination of computer code (as input) and the output from the execution
(be it textual, or visualizations, for example), together with
equations typeset in LaTeX and free-text in one document helps the thinking process. The notebook captures exactly the protocol that was used ({\em i.e.}, order of commands for simulation and analysis) to achieve a certain result~\cite{Beg_2021}. The ability to re-execute a command or simulation easily (because the relevant
commands are readily available in the document) encourages exploration and
verification, and thus supports a learning process that is driven by
experimentation and exploration~\cite{Granger2021} of the behavior of a complex system.
}

\toappendix{While we have not seen this done by our students, it is also possible to author a
project report within the Jupyter notebook. It is thus possible to transition
gradually from a set of instructions for the simulation to run and data to be
plotted to a report that puts those activities and results in context. We
hypothesize that this may lower the barrier towards starting the report writing. Moreover, as demonstrated in previous reports \cite{Odden_2022}, Jupyter notebooks represent a platform that supports and fosters students' epistemic agency as well as reproducibility of the result~\cite{Beg_2021}.
}

\subsubsection{Zero-install software provision with Binder \label{sec:binder}}

\toappendix{In our study, we have made use of the publicly accessible and free Binder software~\cite{Binder-2018}, which
is part of Project Jupyter\footnote{URL: \url{https://mybinder.org/}}. The Binder software takes the URL of a data repository\footnote{Ubermag
repository \url{https://github.com/ubermag/tutorials} on Github}, scans the repository
for files that specify which software is required, installs this
software --- together with a Jupyter notebook server --- in a (Docker) container
image, starts the container, and connects the notebook server from the container
with the user's browser.
None of the technical steps described above are visible to the user:
After selecting the appropriate
URL\footnote{Ubermag on mybinder.org: \url{https://mybinder.org/v2/gh/ubermag/tutorials/latest}}, it takes a
couple of minutes until the desired notebooks session appears in the
browser. The major benefit for our teaching experience is that students can connect to a
Binder session from their desktop, laptop or mobile device, and access the
computational environment in which to experiment (numerically) from their
browser, which helps lowering the usability barrier.}

The public MyBinder service, which hosts the hardware on which the container is executed, comes with some limitations: For example a notebook
session that is idle ({\em i.e.}, no computation and no user activity) for 10 minutes
will be stopped from the MyBinder site and all changes will be lost. The notebook and other
files can be downloaded before the session is stopped (and later uploaded if a
continuation of the work is desired).
The computing hardware offered by MyBinder is relatively weak (for example at most two CPU cores). 

Despite these limitations, MyBinder has been very useful for our teaching
experience in providing a zero-install computational environment: Most students
have carried out all of their simulation computation on the MyBinder service. The reason
the MyBinder service works well for our projects is that the computation
required for the student exercises is relatively modest and can be completed
within minutes to hours on single-core CPUs. If one wanted to offer the same no-install computational environment for
projects that have more substantial computational demands, the university could
host and run their own Binder service: the BinderHub~\cite{Binder-2018} is designed for
this\footnote{We note that JupyterHub is the part of BinderHub responsible for running the server
(after BinderHub builds the image), and that---given appropriate skill sets---it can be configured to have
additional functionality, such as required user authentication, persistent
storage, or control of one or more software environments that can be launched.}.
However, the skills required to set this up exceed those of most
academics, and help from the local computing or IT team is likely to be
required.

A local install of Ubermag on the student's computer is also possible, and some students have chosen to follow
this path. Once the installation is completed, this is more convenient for ongoing and extended studies.

\subsubsection{Zero-install and zero-hosting with JupyterLite\label{sec:jupyterlite}}

Looking ahead, the just emerging JupyterLite
project\footnote{\url{https://jupyterlite.readthedocs.io}, accessed 5/5/2023} circumvents the shortcomings of the MyBinder service. JupyterLite makes
it possible to execute Jupyter Notebooks and many Python packages in the user's
browser (using WebAssembly) and holds great potential for use of software in the
classroom in the future: (i)~like Binder, JupyterLite is a zero-install
approach, and (ii)~the JupyterLite approach does not need other
centralized computing
resources ({\em i.e.}, it is a zero-hosting approach).

In the JupyterLite set up, the complete and pre-configured software environment is
provided for the learners on a (static) web page. Once the webpage is opened by
the learner, the software environment is executed
in the browser of the learner's own device
(computer, laptop, chromebook, ...) which provides the computing power. Such consumer
devices are generally powerful enough, have no limit in run-time, and there is
no dependency on cloud-hosted or other compute resources. (At the moment, the
micromagnetic simulation software is not available as WebAssembly.)

\subsection{User interface design for simulation software in education\label{sec:ubermag-design}}

Our hypothesis is that the use of simulation packages in advanced STEM
classes can have educational value. We had a positive experience using the Ubermag
software. In this section, we describe two important aspects of the user
  interface design. 
\subsubsection{Expose concepts of computational modelling}\label{sec:conc-simul-driv}

When a computer simulation is used to study a science or engineering problem,
there are multiple layers of decision making and simplifications of the problem
taking place (we assume that the model equations include differential equations):
\begin{enumerate}
\item Decide on the model to be used, and express the model in equations.
\item Discretize the model in some form (on grid).
\item Solve the discretized equation.
\end{enumerate}
Many simulation packages are written for a particular model description and
provide all the steps 1 to 3. In particular, the separation between these
different aspects is not visible to the user. In Ubermag, this separation of concerns is more
clearly exposed and accessible, and thus the meaning of the individual steps is
easier for the learner to understand:

\begin{enumerate}
\item Decide on physics approximations and the model to be used: Within the
  Ubermag framework, the user selects the relevant physics through the terms
  that contribute to the energy and dynamics of the system from the
  \texttt{micromagneticmodel} Python package. This creates a
  \emph{machine-readable} definition\footnote{The machine-readable problem
    definition means that a computer (or researcher, educator or learner) can
    read it and extract all needed information to fully define the physics
    problem of interest. For the learning context, the machine-readability
    ensures completeness of information. In a research and industry context,
    machine-readability is a pre-requisite for increasing automation of
    simulation-based work. It also supports reproducibility.} of a micromagnetic problem. Computer scientists
  would express this so that the \texttt{micromagneticmodel} Python package
  provides a \emph{Domain Specific Language} for micromagnetic models of the
  real world~\cite{Beg_2022}.

\item Discretize the model in some form: This requires splitting space into smaller
   parts such as cuboidal cells for finite difference discretization and a wider
   choice of geometrical objects for finite elements. Within the Ubermag
   framework, the \texttt{discretisedfield} package is used to define a (finite
   difference) discretization of space, and scalar and vector fields defined on
   that discretized space.

\item Using the micromagnetic model definition together with the
   discretization, the problem can be solved numerically. Ubermag translates the
   information from the micromagnetic model and the discretized field into a
   configuration file that is understood by one of the micromagnetic
   calculators that it supports.  Using the OOMMF
   Calculator (\texttt{oommfc}) or the mumax$^3$ Calculator (\texttt{mumax3c})
   Python package, Ubermag then delegates the actual numerical solution to OOMMF
   or mumax$^3$.
\end{enumerate}

Through the use of different packages --- with clearly defined and orthogonal
concerns --- the concepts of computational science become easier to grasp
than if all of those aspects are grouped together in the black-box simulation
software.

\subsubsection{Focus on physics, not the package-specific syntax}

A potential user of the software needs to learn and understand what physics
model choice and discretization is implemented in the software, and needs to
learn how to instruct the software (often through a configuration file) to use the
right model, and to combine this with the required geometry, material and other
parameters, initial configuration, time-dependent or spatially-resolved external
effects, etc. Generally, the required configuration file syntax is simulation
package dependent: A scientist (or student) thus needs to learn this syntax for
every new simulation package they want to use, which contributes to the \emph{usability barrier}.

The Ubermag framework provides an abstraction from the specific simulation
package configuration file syntax in the domain of micromagnetics. The
\texttt{micromagneticmodel} package provides the machine readable definition of
the problem using a syntax that scientists perceive as somewhat
intuitive, and Ubermag can automatically translate this into the package-specific configuration file syntax. It is thus much easier to define a
micromagnetic problem with Ubermag than it would be if the packages
OOMMF and mumax$^3$ were used directly. We believe this reduction
in complexity (of specifying a problem in a particular syntax) makes it possible
to explore a much wider set of topics within the teaching module and the computational
projects. This idea --- to provide more ``user-friendly'' high-level interface to
existing simulation software --- is certainly transferable to other domains.
Examples include the atomic simulation environment (ASE)~\cite{Larsen_2017}, and the material
science workflow tool AiiDA~\cite{AiiDA2020}.

\section{Conclusions} \label{sec:conclusion}

We have introduced computer simulation into a materials science class,
and describe approaches we found beneficial for the
learning experience. It would be interesting to evaluate these in the context of
different subject areas and educational settings:
\begin{itemize}
\item Choice of Python as one language for simulation and analysis, with broad
library support.

\item Design and use of user interfaces that focus on the learner's
  interest: expose modelling concepts and hide peculiarities and complexity of
  underlying simulation engines.

\item Use of Jupyter Notebooks to encourage interactive exploration of the
(simulated) system under study.

\item Binder capabilities of project Jupyter which make it possible to execute
  simulations in the cloud rather than on the students' computers. This
  overcomes the (sometimes significant) challenge of installing the software
  locally.
\end{itemize}

Extensions of work described here include combinations of Jupyter based
simulation teaching with computational essays~\cite{DiSessa_2000,Odden_2022}, the
``nbgrader''' tool supporting the grading in Jupyter notebooks~\cite{nbgrader2019},
and the opportunities for the JupyterLite (Sec.~\ref{sec:jupyterlite}) based
zero-install zero-hosting provisioning of software at scale.

\section{Supplementary Materials}
Please click on this link to access the supplementary material, which includes a detailed overview of five simulation projects, the problem sheets, a general overview of computational methods in education, a thematic analysis of student and teacher feedback on the computational projects, and additional practical considerations. Print readers can see the supplementary material at [DOI to be inserted by AIPP].

\section*{Acknowledgements}
\censorbox{
The authors would like to thank Thomas Wilhelm (Goethe University Frankfurt) for
fruitful discussions on numerical methods in higher education, and Min Ragan-Kelley for helpful discussions on project Jupyter. MLo acknowledges
the financial support by the German Science Foundation (Deutsche
Forschungsgemeinschaft, DFG) through the research fellowship LO 2584/1-1 for the development of the simulation projects 1 and 2, the implementation of student interviews and questionnaires, as well as the manuscript preparation. SH, SP
and HF were supported by the Engineering and Physical Sciences Research
Council’s United Kingdom Skyrmion Programme Grant (EP/N032128/1). The
development of the MSE 598/498/464 course and specifically the simulation
projects 3 and 4, as well as the efforts from RK and AH were partially supported
by the NSF through the University of Illinois at Urbana Champaign Materials
Research Science and Engineering Center Grant No. DMR-1720633. The development
of the simulation project 5 by T-HL was supported by the US Department of
Energy, Office of Science, Materials Science and Engineering Division, under
Contract No. DE-SC0022060.
}
The authors have no conflicts to disclose.

\bibliography{NumSim}


\end{document}



\title{Supplementary material for \\ ``Numerical simulation projects in micromagnetics with Jupyter"}

\author{Martin Lonsky}
\email{lonskymartin@gmail.com}
\affiliation{Materials Research Laboratory and Department of Materials Science and Engineering, University of Illinois at Urbana-Champaign, Urbana, Illinois 61801, USA}
\affiliation{Institute of Physics, Goethe University Frankfurt, 60438 Frankfurt, Germany}
\author{Martin Lang}
\affiliation{Max Planck Institute for the Structure and Dynamics of Matter,
  22761 Hamburg, Germany}
\affiliation{Faculty of Engineering and Physical Sciences, University of Southampton, Southampton, SO17 1BJ, United Kingdom}
\author{Samuel Holt}
\affiliation{Faculty of Engineering and Physical Sciences, University of Southampton, Southampton, SO17 1BJ, United Kingdom}
\affiliation{Max Planck Institute for the Structure and Dynamics of Matter, 22761 Hamburg, Germany}
\author{Swapneel Amit Pathak}
\affiliation{Faculty of Engineering and Physical Sciences, University of Southampton, Southampton, SO17 1BJ, United Kingdom}
\affiliation{Max Planck Institute for the Structure and Dynamics of Matter, 22761 Hamburg, Germany}
\author{Robin Klause}
\affiliation{Materials Research Laboratory and Department of Materials Science and Engineering, University of Illinois at Urbana-Champaign, Urbana, Illinois 61801, USA}
\author{Tzu-Hsiang Lo}
\affiliation{Materials Research Laboratory and Department of Materials Science and Engineering, University of Illinois at Urbana-Champaign, Urbana, Illinois 61801, USA}
\author{Marijan Beg}
\affiliation{Department of Earth Science and Engineering, Imperial College London, London SW7 2AZ, United Kingdom
}
\author{Axel Hoffmann}
\affiliation{Materials Research Laboratory and Department of Materials Science and Engineering, University of Illinois at Urbana-Champaign, Urbana, Illinois 61801, USA}
\author{Hans Fangohr}
\email{hans.fangohr@mpsd.mpg.de}
\affiliation{Max Planck Institute for the Structure and Dynamics of Matter, 22761 Hamburg, Germany}
\affiliation{Faculty of Engineering and Physical Sciences, University of Southampton, Southampton, SO17 1BJ, United Kingdom}
\affiliation{Center for Free-Electron Laser Science, 22761 Hamburg, Germany}

\date{\today}

\begin{abstract}
  The present document contains supplementary information for the article ``Numerical simulation projects in micromagnetics with Jupyter." This involves a general overview of computational methods in education (Sec.~\ref{Overview}), a detailed description of five simulation projects (Sec.~\ref{Projects}), additional practical considerations (Sec.~\ref{supplement:practical-considerations}), student and teacher feedback on the learning experience (Sec.~\ref{sec:feedback-learning-experience}), and the computational problem sheets (Appendix).
\end{abstract}

\maketitle

\tableofcontents

\section{Incorporation of Computational Methods into STEM Programs} \label{Overview}
In this section, we review the current status of teaching computational
modeling and numerical simulations at university level. This involves several
examples of already existing classes and tools, as well as suggestions for
further applications in undergraduate courses. 

There are various ways how computational methods can be embedded into
traditional courses that form an existing undergraduate or graduate curriculum.
In general, this may be done in the form of computational assignments,
computational lectures, computational laboratories, and computational research
projects~\cite{Kononov_2017}. In the case of physics, an excellent platform for
relevant course materials, teaching approaches, workshops, and interaction with
fellow teachers has been established by the Partnership for Integration of
Computation into Undergraduate Physics (PICUP)
organization~\cite{Caballero_2019}. A further comprehensive compilation of useful resources on computational physics can be found in Ref.~\cite{Atherton_2023}.
In the remainder of this section, we will give several
examples of how to embed computational methods into STEM programs.

Computational activities can be incorporated as exercises into introductory courses, see for
example Ref.~\cite{Kortemeyer_2018}. In laboratory courses,
experiments can be combined with computer simulations, for example by using
Visual Python (VPython)~\cite{Caballero_2012}, Mathematica~\cite{Samsonau_2018} or 
Microsoft Excel~\cite{Sachmpazidi_2021}. VPython is a visual extension of Python that can be utilized for simulating and visualizing simple experiments with only a few lines of code~\cite{Kortemeyer_2020}. For
instance, VPython has been used in the aerospace engineering program at the
University of Southampton, where second-year students were assigned
computational projects related to three-dimensional real time
visualization~\cite{Fangohr_2006}.
In physics, VPython is particularly
useful for classical mechanics classes and can nowadays also be run in Jupyter
Notebooks and in browsers by using Web VPython.
VPython can also be presented with an introduction to programming if students have no prior programming experience~\cite{Fangohr_2006}. In terms of closed computational environments,
Physlets~\cite{Christian_1998}
and the Physics Education Technology (PhET) Interactive
Simulations project at the University of Colorado Boulder~\cite{Perkins_2006}
are noteworthy products for introductory physics courses. 

A major portion of
open computational environments is focused on more advanced and specialized
topics, such as the interactive molecular dynamics simulation code developed at
Weber State University~\cite{Schroeder_2015}, and the nanoHUB platform~\cite{Madhavan_2013} developed at Purdue University. The latter offers various accessible simulation software packages for semiconductor and nanotechnology applications. At {the University of Illinois at Urbana-Champaign (UIUC)},
instructors have started using some of the resources for various undergraduate
courses~\cite{Mansbach_2016, Mansbach_2016b, Kononov_2017, Zhang_2018,
  Lee_2019}. This involves several homework assignments where students have to solve problems using computational methods. For instance, in the ``Electronic Properties of Materials'' (MSE
304) course, students use
density functional theory (DFT) to investigate properties such as the density of
states, band structure and effective masses of electrons and holes in different
materials by utilizing the Quantum ESPRESSO computer
code~\cite{Giannozzi_2009,
  Giannozzi_2017} on nanoHUB. In another
assignment, students are asked to use ABACUS (Assembly of Basic
Applications for the Coordinated Understanding of Semiconductors)~\cite{Klimeck_2009} to obtain a qualitative and quantitative understanding of a
semiconducting diode. Aside from the nanoHUB toolset, a comprehensive
overview of relevant methods in the context of teaching computational materials
science and engineering can be found in Ref.~\cite{Thornton_2009}.

The inclusion of computational methods in existing
courses will also benefit from simulation environments such as
Ubermag~\cite{Beg_2022} and the Atomistic Simulation Environment
(ASE)~\cite{Larsen_2017}.  ASE is a software package for atomistic
simulations provided by the Technical University of Denmark.
Ubermag is described in more detail in Sec.~II of the main article.
Both Ubermag and ASE provide a high-level (i.e., simplified and accessible) interface to
multiple simulation packages: Controlling these simulation packages directly
would require more specialized skills than using the high-level interface. 
Both ASE and Ubermag use a Python interface and are supported by an
extensive documentation, which includes tutorials, frequently asked questions,
and contents of workshops. 
These simulation environments offer a \emph{lowering of the usability barrier} which is essential for educational processes. Students can change the simulations inputs and understand how the outputs change in response without having to learn the fine details of the syntax. This is more
efficient if the changes of the input are concise and cognitively not too
demanding. Other examples of software with high-level interfaces and thus good potential
for teaching include COMSOL
Multiphysics~\cite{Pieper_2014} and the Einstein Toolkit
which offers computational tools for relativistic astrophysics and gravitational
physics applications~\cite{Loeffler_2012}.

A further interesting method to embed
computational contents into the curriculum is given by so-called computational
essays~\cite{Odden_2022}. Originally proposed in a work by diSessa~\cite{DiSessa_2000}, computational essays consist of a combination of text,
executable code, interactive diagrams and other computational tools. As pointed out by Odden \textit{et al.}, computational essays can help students to develop epistemic agency, i.e., a greater control over their own learning process \cite{Odden_2022}. For the
realization of such essays, interactive Jupyter notebooks~\cite{Granger2021} represent an ideal
tool. 

Following this overview of previous efforts on incorporating computation into STEM curricula, we now turn to the detailed description of simulation projects that we used in our course on magnetic materials and applications.

\section{Description of Simulation Projects} \label{Projects}

This section includes a description of the micromagnetic simulation projects that we developed at {the University of Illinois at Urbana-Champaign}, which may be particularly useful for instructors in the field of magnetism, but it can also serve as an inspiration for teachers in other domains. Note that the complete problem sheets are appended at the end of this document.

\subsection*{Project 1: Static and dynamic properties of a magnetic vortex} 
Ferromagnetic disks with lateral sizes in the micrometer or submicrometer range
exhibit under certain conditions a configuration that is termed magnetic vortex~\cite{Shinjo2000}.
In the first subproject, the students are asked to determine the minimized
energy state of a well-defined nanodisk geometry, which corresponds to such a
vortex state. Subsequently, the students play around with various parameters in
order to discuss the chirality and polarity of magnetic vortices, the dependence
on the magnetic anisotropy of the material, and the changes as a function of
external magnetic field. The second subproject appears significantly shorter,
but in reality it is more complex and challenging -- a fact that is
unambiguously stated on the problem sheet for the sake of transparency. In other
words, by offering tasks with varying degrees of difficulty we can cater to a
broad spectrum of students with different backgrounds.

In this case, the students are asked to simulate the nucleation and annihilation
of a magnetic vortex by sweeping the magnetic field and thus determining the
magnetic hysteresis curve. Moreover, they have to obtain snapshots of the
magnetization such that they are able to understand the different mechanisms (e.g., domain wall motion, rotation of magnetization, etc.) that are responsible for certain features in the hysteresis loop. Finally, in
the third subproject, the students investigate gyrotropic dynamics of the
magnetic vortex by applying a magnetic field pulse which leads to a periodic
motion of the spin texture. This task involves the calculation of a fast Fourier
transform in order to obtain the power spectrum and thus the resonance peak at a
specific frequency. 

\begin{figure*}
  \centering
  \includegraphics[width=10cm]{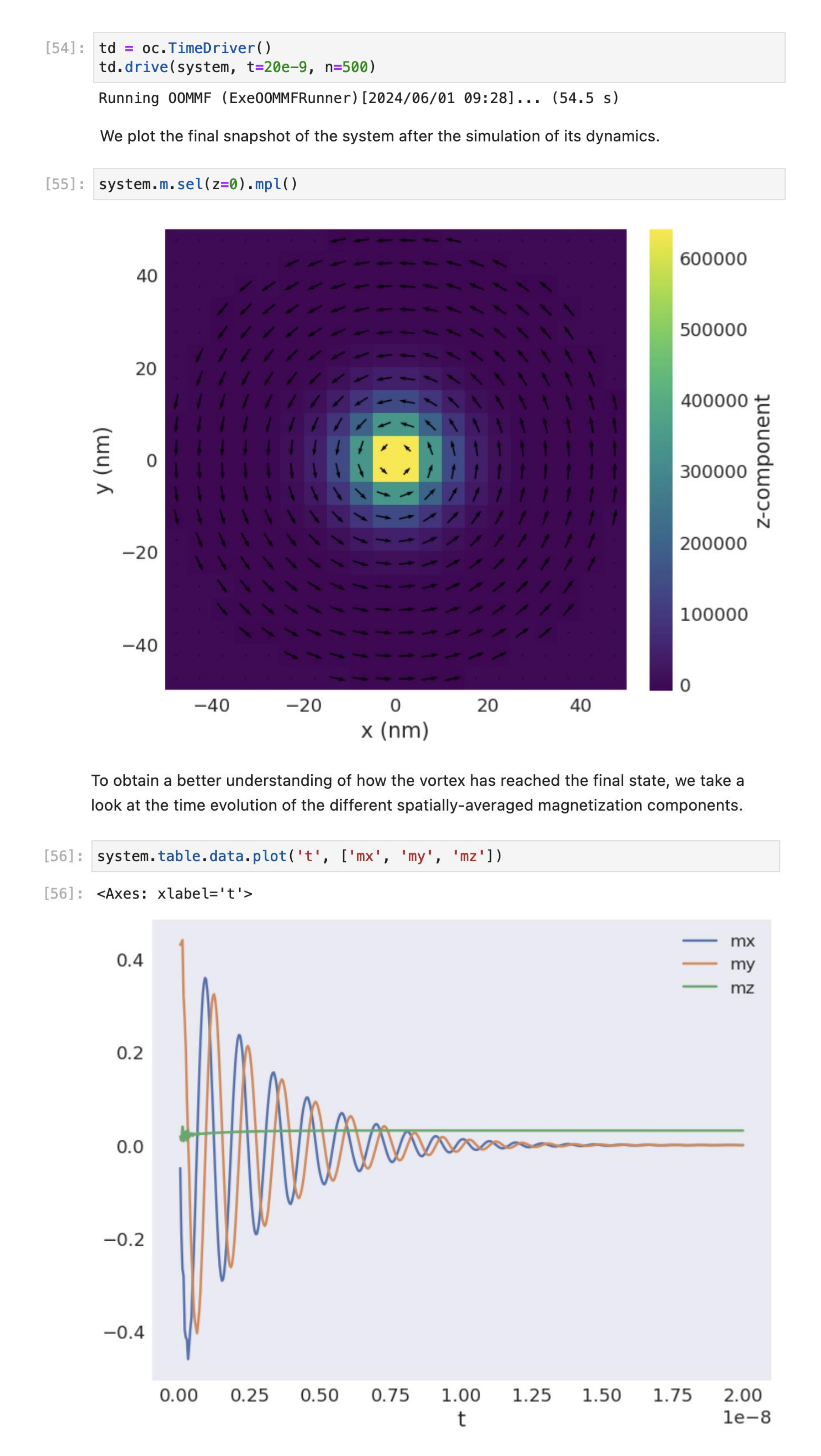}
  \caption{Screenshot of a Jupyter notebook containing Ubermag-based micromagnetic simulations used in a web browser via Binder. This excerpt contains Python code, Markdown text and two diagrams. See main text for details.} 
  \label{EXAMPLE_JUPYTER}%
\end{figure*}

A screenshot of a Jupyter notebook containing calculations
that are related to this subproject is depicted in Fig.~\ref{EXAMPLE_JUPYTER}. This particular notebook contains Python
code, Markdown text, an image to visualize the magnetic configuration of a
nanodisk with a diameter $d=100\,$nm, and a plot of the time-dependence of
the (spatially averaged) magnetization components $m_\mathrm{x}$, $m_\mathrm{y}$ and $m_\mathrm{z}$ after excitation by a magnetic field
pulse. All simulation projects in our course are carefully designed such that each
calculation runs for a reasonably short period of time, i.e., seconds to
minutes. 

\subsection*{Project 2: Synthetic antiferromagnets}
Synthetic antiferromagnets~\cite{Duine2018} are typically composed of two ferromagnetic layers
that are antiferromagnetically coupled via the so-called
Ruderman-Kittel-Kasuya-Yosida (RKKY) interaction~\cite{Parkin1991} through a nonmagnetic spacer.
More specifically, the antiferromagnetic interlayer exchange interaction is
mediated by the conduction electrons in the spacer layer. In the first subtask,
students explore the role of different micromagnetic energy terms with regard to
the minimized energy state of the synthetic antiferromagnet. For instance, the
dipolar interactions (demagnetization energy term) are neglected initially and
the system is relaxed from an initial random magnetization state.
Subsequently, dipolar interactions are added to the Hamiltonian of the system
and the result is compared to the previous simulations. In the next step, the
influence of perpendicular anisotropy and Dzyaloshinskii-Moriya interaction (DMI) is investigated. In the second
subproject, skyrmions in a synthetic antiferromagnet with DMI are studied. To
this end, the students have to create a skyrmion in the simulated geometry and
then study its properties as a function of varying micromagnetic energy terms.
Students should also explore the possibility of creating multiple skyrmions in
the nanostructure. In the third subproject, the magnetization reversal of a
synthetic antiferromagnet with around 20 layers is studied. This is the most challenging
part, since a hysteresis loop of a complex magnetic structure needs to be
simulated. Eventually, students can observe the switching behavior in the
individual layers of the structure by looking at snapshots.

\subsection*{Project 3: Static and dynamic properties of domain walls}
Domain walls are present in ferromagnetic materials in order to reduce the
magnetostatic energy of the system. Their nucleation and propagation is one of
the magnetization reversal mechanisms. This project allows the students to study
domain wall initialization in systems with different energy configurations, the
motion of domain walls across a magnetic strip, the Walker breakdown~\cite{Schryer1974}, and the
pinning of a domain wall to a notch in a magnetic strip. While the first
subproject revolves around the static properties of a domain wall in a magnetic
strip, the second subproject focuses on the domain wall Walker breakdown.

\subsection*{Project 4: Static and dynamic properties of vortices in an elliptical geometry}
This is a further project on magnetic vortices, however, it is centered around
the role of sample geometry. In particular, an elliptical sample is considered,
whereby the major and minor axis lengths are varied in order to study the
changes in the magnetic configuration accordingly.

\subsection*{Project 5: Ferromagnetic resonance of a Pt/Co nanodisk} 
Ferromagnetic resonance is an excitation of a magnetic material during which all
moments exhibit an in-phase precession. It represents a unique spectroscopic
technique~\cite{Mewes2021} to investigate the dynamic properties of a system and to extract
essential quantities such as the Gilbert damping. In this project, the dynamic
properties of a well-defined Pt/Co nanodisk are investigated by the students.
This implies the simulation of the ferromagnetic resonance as a function of
different external magnetic fields and the calculation of the corresponding
resonance frequencies in the first subproject. Furthermore, the role of the
Gilbert damping is also studied. In the second subproject, students investigate
the impact of the size of the system and the direction of the external magnetic
field on the dynamic properties.

\section{Additional practical
  considerations} \label{supplement:practical-considerations}

In the following, we provide additional practical considerations about open source software and an additional Ubermag package that connects micromagnetic simulations with experimental results.

\subsubsection{Open source software}

Ubermag and the software from Project Jupyter (such as the Jupyter Notebook and
Binder) are freely and openly available as open-source software.
In contrast to commercial simulation packages such as COMSOL Multiphysics~\cite{Pieper_2014,
  Dedulle_2007} and other, Ubermag and
Python-based Jupyter notebooks offer enhanced value, transparency and
unlimited accessibility for students, instructors, and departments~\cite{UbermagSource}. There is also no tie-in to a particular vendor: Should the university dislike the way the Ubermag package develops in the future, it could take the current
Ubermag version and either keep using it as is, or modify it to suit its own
needs best.

The ideal scenario is of course that users of the package feed back any requests
for improvements (or actual code changes that implement these improvements) to
the open source team. There are well established protocols for such
contributions, although it may need skills beyond those of the average academic.
Increasingly, universities employ research software engineers who can provide
such skills~\cite{Cohen2021}.

\subsubsection{Fostering the link between micromagnetic simulations and experiments}

A relatively new package termed
\texttt{mag2exp} of Ubermag~\cite{Holt_2023}
allows to simulate physical quantities identical to
those obtained from experimental techniques that are used to study magnetism,
such as Lorentz transmission electron microscopy or small angle neutron
scattering. While we have not implemented this in our simulation projects yet,
this approach is expected to further reduce the gap between experiments and
numerical modeling and thus represents a possible extension of the existing
exercises. 

\section{Feedback on Learning
  Experience}\label{sec:feedback-learning-experience}

\subsection{Research Questions}
We collected feedback on the learning experience from both students and the teaching staff with the aim of addressing the following two research questions.

\textbf{(RQ1)} What are the critical factors and key challenges to consider for a successful and effective incorporation of computational methods into an existing course? 

\textbf{(RQ2)} What are the benefits and challenges perceived by students in regard to computational projects in a science or engineering class?

Prior to a discussion of the feedback, we describe the methodology of our work. This involves a specification of participants, the instruments for data collection as well as the data analysis procedures. 

\subsection{Methodology}
Personal interview surveys were carried out in October 2022 with six students of the 2020 and 2021 classes in order to assess their experience with the simulation
projects~\footnote{Prior to the interviews, we had reached out to 20 former students whose email addresses were still known and received a response by six individuals.}. All of them were Master or PhD students at the time of attending the course and had differing levels of experience in computational methods and programming. 
The conversations lasted about 30 minutes per student. During the personal interview survey we asked
open-ended questions regarding the following points:

\begin{itemize}
\item Overall experience with simulation projects, suggestions for improvements.
\item Difficulty of projects, taking into account students' proficiency in
  programming and their prior experience in the field
  of magnetism.
\item Assessment of the importance of computational methods for research and
  whether micromagnetic simulation projects changed students' attitude.
\item Communication and collaboration within the student groups as well as the
  communication between the students and the instructional staff.
\end{itemize}

In addition, we also collected feedback from three teaching assistants (two PhD students and one postdoc) and from the main instructor (full professor) of the class through personal interviews. In the 2020 and 2021 courses, there were two teaching assistants responsible for supervising the projects. Although a single teaching assistant could have managed the workload, we chose to increase the staff to provide more personalized attention for the students.      
After completing data collection, we performed a thematic analysis by following the scheme recommended by Braun and Clarke \cite{Braun_2006}. In the following, we present our analysis which is divided into two parts, taking into account feedback from students as well as instructors. 

\subsection{Feedback from Students}
In the context of student feedback, we have identified six themes that are relevant to the above-mentioned research questions, which we classify into positive and negative themes. An overview of the themes along with a brief definition of each theme is provided in Table \ref{Tab:Students}.

\begin{table*}
\centering
\caption{Themes from students' response provided in personal interviews and questionnaires.}
\begin{ruledtabular}
\begin{tabular}{lp{110mm}}
\multicolumn{1}{c}{Themes} & \multicolumn{1}{c}{Definition}  \\
\hline
\textit{Positive themes}  &           \\
Physical intuition and epistemic agency & Freedom to explore by varying micromagnetic energy terms; better understanding through immediate visualization; self-paced learning. \\
Easy accessibility of simulations & Using Ubermag in the cloud, no installation required; obtain experimentalist's view on magnetism without need for expensive infrastructure.  \\
Effective feedback loops and communication & Intermediate discussions with teaching staff can lead to better results; mostly good communication within student groups using different media.   \\
Relevance for academia and industry & Understand importance of computation in modern research and industry; programming with Python is a valuable skill and experience with numerical simulations can be beneficial for future career.   \\
\hline
\textit{Negative themes}  &                                  \\
Varying levels of experience and skills & Challenges originating from diverse background of undergraduate and graduate students; especially programming skills vary strongly.  \\
Technical details and limitations & Artifacts in simulations, validity of numerical methods and geometric constraints (e.g., edge effects) should be discussed in more detail.                                
\end{tabular}
\end{ruledtabular}
\label{Tab:Students}
\end{table*}

\subsubsection{Physical intuition and epistemic agency}

In the personal interviews, four out of six students explicitly stated that varying the different micromagnetic energy terms
and visualizing the magnetization states has led to a better understanding of the topic in comparison to problem sheets or demonstrations in the lecture. We observed that students develop epistemic agency through self-paced working. Such an in-depth exploration of the complex relationships in micromagnetics cannot be conveyed easily in a conventional lecture format with limitations such as time constraints. One student stated that their physical intuition for magnetism has significantly improved by working on these projects. As shown in Subsec.~II~B~of the main article, it is the complexity of multiple competing energy terms in micromagnetics that makes the simulations a rich and instructive playground. We emphasize that some students even explored topics that were beyond the scope of the required tasks.

\subsubsection{Easy accessibility of simulations}

Two of the interviewed students noted that the simulations offered an experimentalist's perspective on magnetism without the need of being in a laboratory environment. Even though this does not imply that computational methods should replace experimental lab courses, they certainly represent a valuable and accessible addition that does not require extensive resources. In the context of accessibility, we also highlight the positive feedback on the Ubermag software as a well-suited teaching tool in comparison to related packages such as OOMMF and mumax$^3$. Its intuitive handling and straightforward visualization capabilities were particularly appreciated by students with limited background in computational methods. Moreover, the fact that Ubermag can be executed in the cloud free of charge and without installation on the students' personal computer enhanced the accessibility. More details along with practical recommendations for computational projects can be found in Sec.~III of the main article.

\subsubsection{Effective feedback loops and communication}

The communication within the project groups was typically perceived as positive and took place in the form of virtual meetings, emails or text messages. The collaboration was viewed as fruitful and effective in a majority of groups, since students reportedly exchanged ideas, comments and concerns about the different subtasks. The communication between the students and the teaching staff was also assessed positively. It was pointed out by several students that the intermediate discussions were particularly valuable to them because of important preliminary feedback and help with potential issues. 

\subsubsection{Relevance for academia and industry}

Each of the six interviewed students recognized that computational methods are important in today's research landscape. One student shared that they had been intimidated by computational methods prior to taking the course, but that the simulation projects resulted in a more positive mindset and a better understanding of the underlying concepts. Some students have learned to recognize the importance of computational methods in science after working on the projects, while others viewed the simulations as relatively limited in terms of their scope and duration, and thus their attitude has not changed significantly. 
We note that the Department of Materials Science and Engineering at UIUC has already incorporated computational methods into several core classes of the undergraduate curriculum and therefore students in this program likely have a clearer picture with regard to the importance of computational methods in research than students at other institutions. Consequently, working on the micromagnetic simulation projects may not have significantly impacted the attitude of those students who completed their undergraduate degree at the same institution, compared to those who came to UIUC after attending college elsewhere.  
Students also demonstrated a clear understanding of the relevance of computational methods with respect to their future professional career. Interestingly, the simulation projects as well as the gained experience in programming with Python became a central topic in one student's job interview and appeared to be assessed as positive by the interviewers. 

\subsubsection{Varying levels of experience and skills}

Prior to working on the simulation projects, most students indicated that they were only barely or somewhat familiar with numerical methods, but on average reasonably comfortable with
programming, especially with Python. Although the limited experience with numerical methods seems contradictory to the department's strong focus on computational methods in undergraduate education, this can be explained by the fact that most graduate students in our course came from external institutions with potentially weaker backgrounds in computation.  
A majority of students claimed to have a very limited experience with magnetism.
The fact that the course is attended by both undergraduate and graduate students poses a challenge for the instructor to guarantee an effective learning experience for everybody. However, despite their different background knowledge, nearly all students stated that the difficulty of the projects was appropriate. The extensive
Ubermag documentation and collection of example notebooks have been very beneficial to address the variety of existing skills.
A majority of students pointed out that the projects were more enjoyable than regular problem sheets assigned in the class.  

\subsubsection{Technical details and limitations}

The student feedback suggests that a more comprehensive introduction to Ubermag going beyond the brief presentation and materials that we provided will be beneficial in the future. In particular, an additional session with short exercises prior to the assignment of projects could be offered by teaching assistants in future courses. Moreover, the capabilities, validity and limitations of computational methods could be discussed in more detail. One example is the role of possible artifacts that are caused by an inappropriately-chosen mesh shape or size.

\subsection{Feedback from Teaching Staff}

Based on the feedback from teaching staff members, we have identified three themes. We present these themes together with a brief definition in Table \ref{Tab:Educators}.

\begin{table*}
\centering
\caption{Themes from teaching staff's response provided in personal interviews.}
\begin{ruledtabular}
\begin{tabular}{lp{110mm}}
\multicolumn{1}{c}{Themes} & \multicolumn{1}{c}{Connection to research question}  \\
\hline
Project development and work load   &   Time-consuming at first, but projects can be reused each year. \\
Communication and guidance & Good communication between students and teachers even during pandemic. \\
Modern topics and alignment with lecture & Selecting appropriate topics for projects allows students to work on modern research questions. Difficult to achieve alignment with rest of the class.   \\
                              
\end{tabular}
\end{ruledtabular}
\label{Tab:Educators}
\end{table*}

\subsubsection{Project development and work load}

Initially, the design of the computational projects from scratch was laborious and time-consuming for the teaching assistants, especially due to the required careful testing of example solutions. However, the problem sheets and solutions could be reused in the subsequent semesters while gradually fixing issues that were discovered. 
Apart from designing the projects, leading intermediate discussions and moderating the session with the students' final presentations, the feedback from the teaching staff indicates that their work load was not exceedingly high compared to other courses. 

\subsubsection{Communication and guidance}

Discussions and intermediate meetings with students were helpful for the teaching staff to identify mistakes and ambiguities in the problem definition that had caused students to obtain erroneous simulation results or get lost in irrelevant details. 
Despite the challenging circumstances in the fall 2020 and 2021 semesters due to the COVID-19 pandemic, the communication between the instructional team and students as well as within the student groups was clearly more intensive than in courses without computational group projects. Students were encouraged to actively discuss about the subject matter and ask questions to their peers as well as to instructors. 

\subsubsection{Modern topics and alignment with lecture}

Overall the simulation projects were designed to relate directly to current
research topics, as well as to distinct topics of the regular lectures. For
example, synthetic antiferromagnets are integral components of modern
spintronics devices to minimize complications from stray fields. Similarly, the
complex gyrotropic nature of magnetization dynamics gives rise to phenomena
such as Walker breakdown for moving domain walls, or distinct chiralities in the
motion of topological solitons such as skyrmions. Unfortunately, given the extended timeframe of the micromagnetic simulation
projects, it was not always possible to align the projects
well with the discussion of these concepts. We consider the use of small example notebooks and shorter homework problems to achieve that alignment in the future. For instance, at the moment, the projects do not include any current-driven dynamics, since spin transfer torques are only discussed in class after the conclusion of the projects. Assigning multiple short projects at different points throughout the semester instead of a single comprehensive project would allow for more flexibility and better alignment with the syllabus. 

We note that the Ubermag numerical simulations fit perfectly into the syllabus,
since micromagnetic modeling has been discussed in our course
prior to the introduction of the simulation projects and is also typically
presented in magnetism courses at other universities. 

\bibliography{NumSim}

\clearpage
\foreach \x in {1,...,3}
{%
\includepdf[pages={\x,{}}]{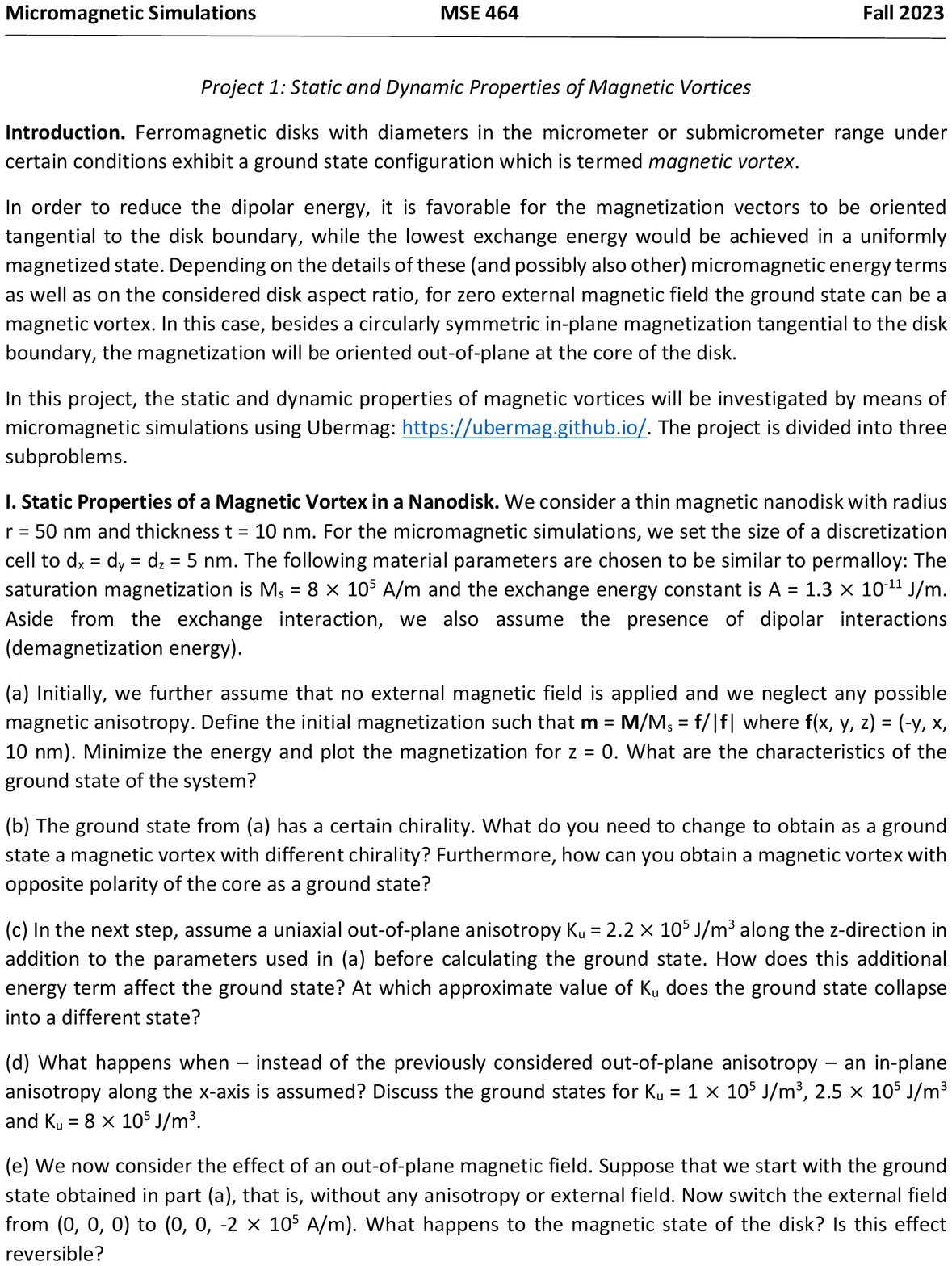}
}
\foreach \x in {1,...,3}
{%
\includepdf[pages={\x,{}}]{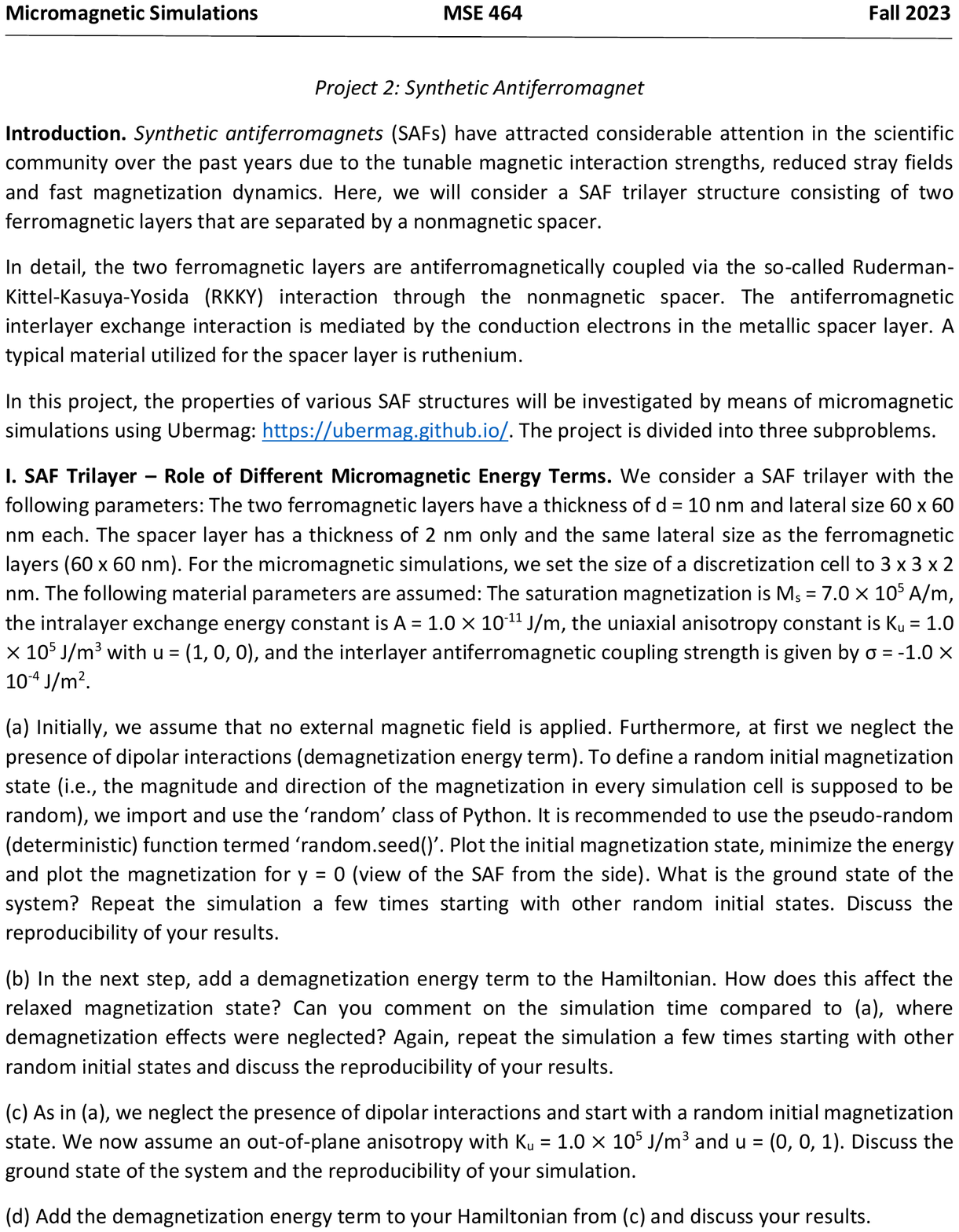}
}
\foreach \x in {1,...,3}
{%
\includepdf[pages={\x,{}}]{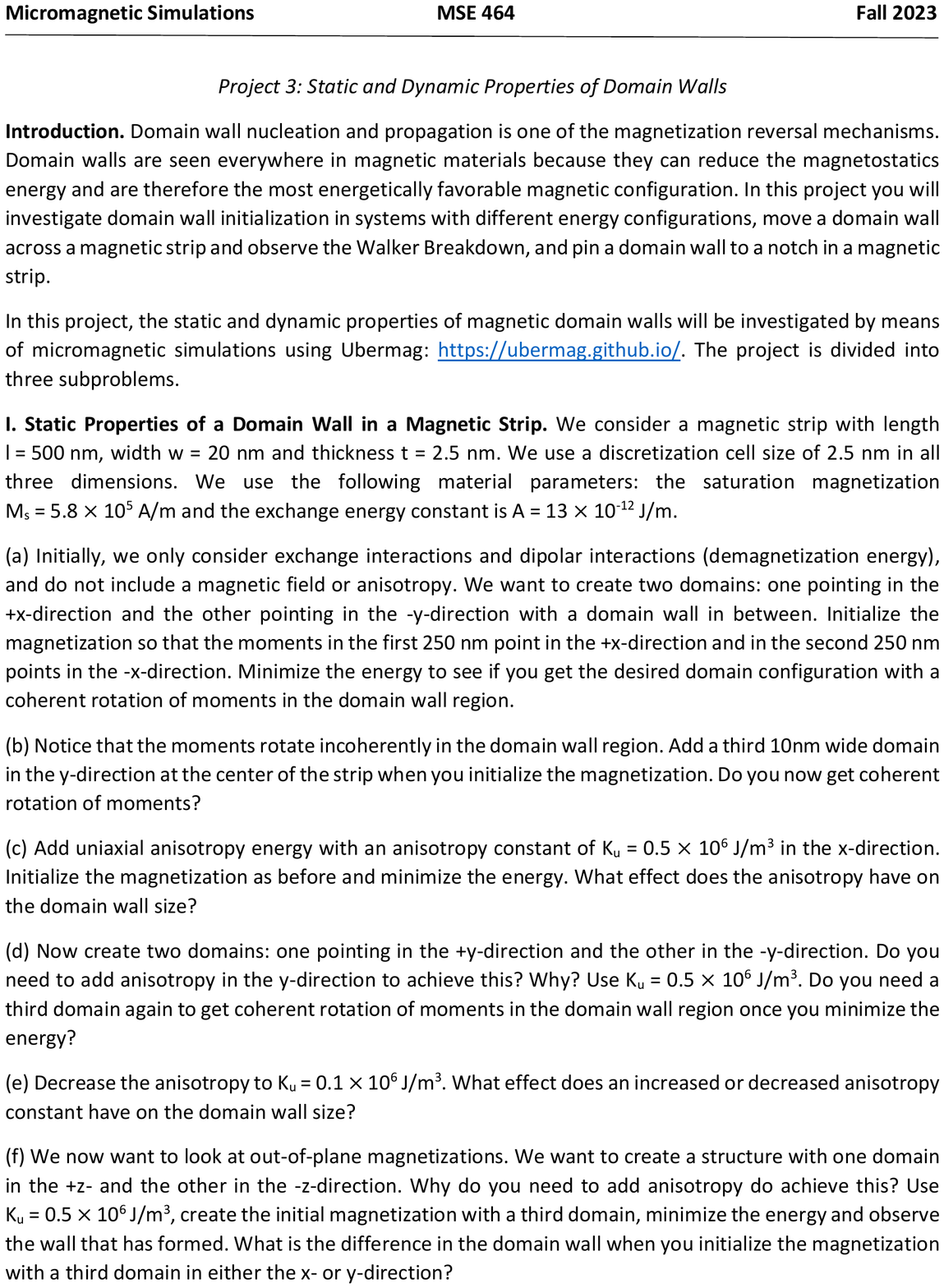}
}
\foreach \x in {1,...,3}
{%
\includepdf[pages={\x,{}}]{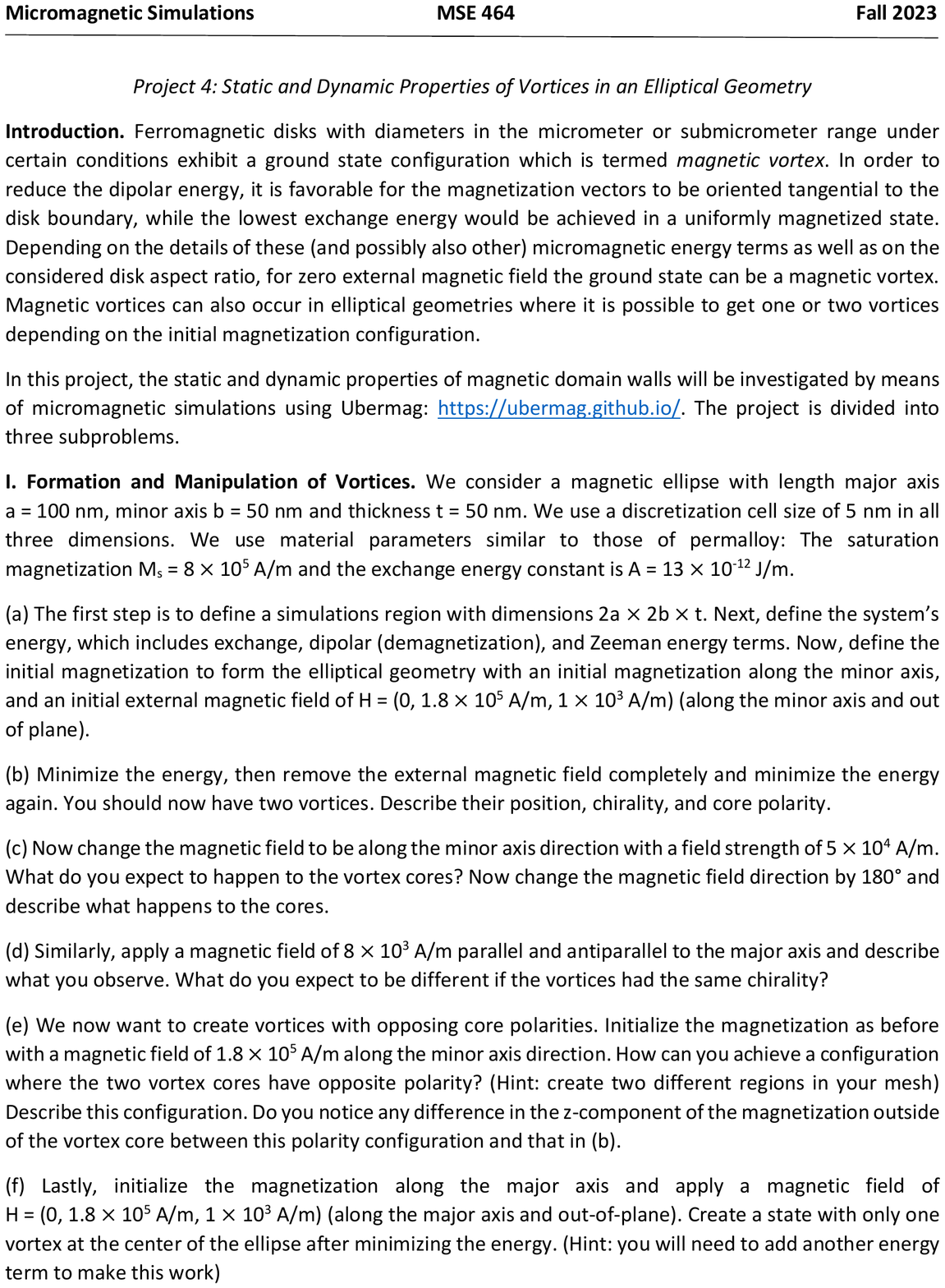}
}
\foreach \x in {1,...,3}
{%
\includepdf[pages={\x,{}}]{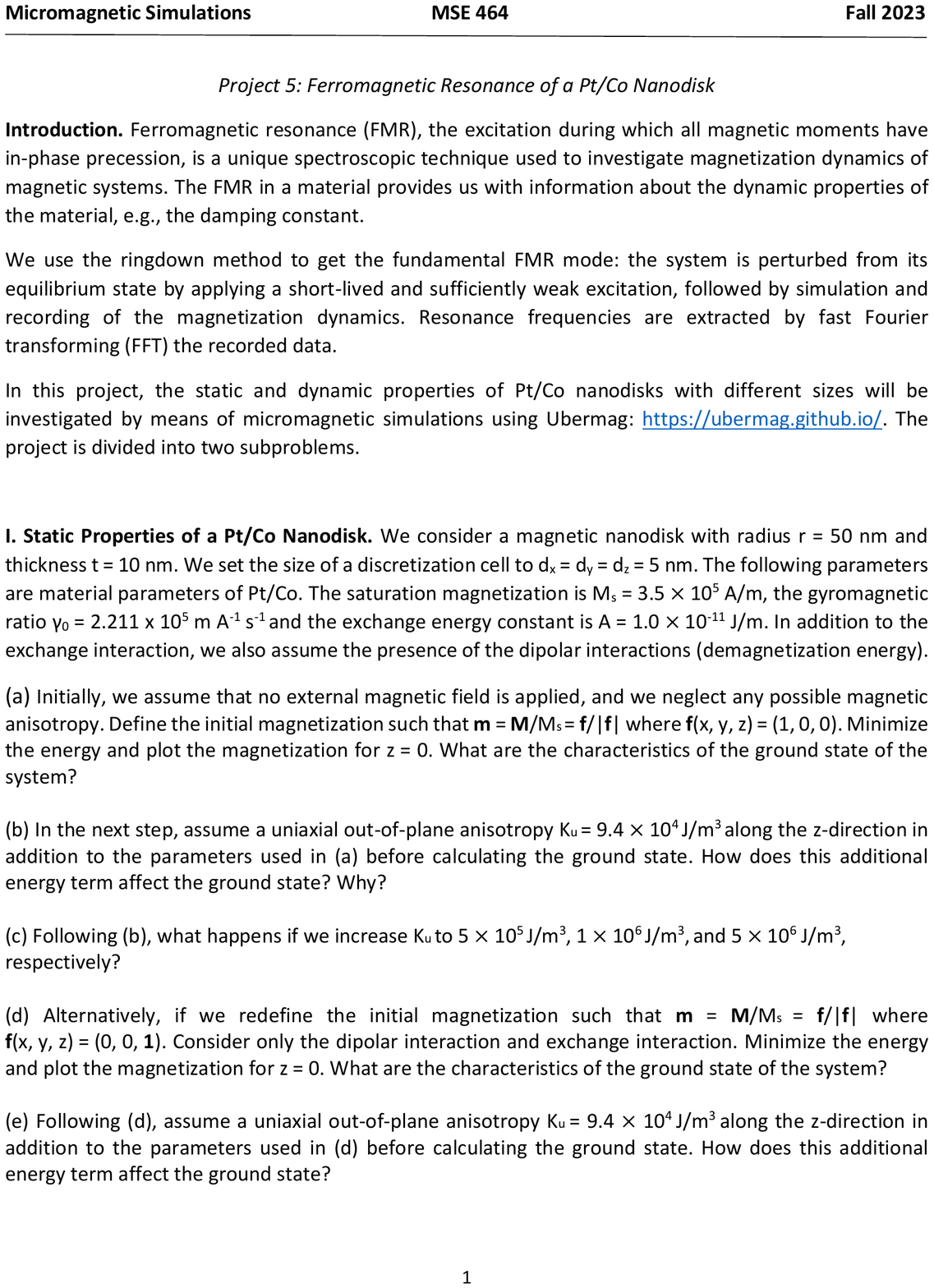}
}